\documentclass[aps,prd,showpacs,amssymb,superscriptaddress, notitlepage,floats,floatfix,nofootinbib]{revtex4-2}

\pdfoutput=1

\usepackage{booktabs}
\usepackage{graphicx}
\usepackage{subfig}
\usepackage{amssymb}
\usepackage{amsmath}
\usepackage{amsfonts}
\usepackage{tensor}
\usepackage{bm}
\usepackage{mathrsfs}
\usepackage[colorlinks=true]{hyperref}
\usepackage[all]{hypcap}
\usepackage[utf8]{inputenc}
\usepackage{color}
\usepackage{multirow}
\usepackage{palatino}
\usepackage{float}
\usepackage{siunitx}

\usepackage{soul}
\usepackage{calrsfs}

\usepackage{graphicx}
\usepackage{xcolor}

\newcommand{\R}[1]{{#1}}

\begin{document}
\title{Non-linear stability of soliton solutions for massive tensor-multi-scalar-theories.}
\author{Riccardo Falcone}
\email{riccardo.falcone@studenti.unimi.it}
\affiliation{Theoretical Astrophysics,  IAAT, University of T\"ubingen, T\"ubingen 72076, Germany}
\affiliation{Dipartimento di Fisica, Universit\`a degli Studi di Milano, Via Celoria 16, 20133, Milano, Italy}, 
\affiliation{Dipartimento di Fisica, ``Sapienza'' University of Rome, Piazzale Aldo Moro 5, 00185, Roma, Italy}

\author{Daniela D. Doneva}
\email{daniela.doneva@uni-tuebingen.de}
\affiliation{Theoretical Astrophysics, IAAT, University of T\"ubingen, T\"ubingen 72076, Germany}

\author{Kostas D. Kokkotas}
\email{kostas.kokkotas@uni-tuebingen.de}
\affiliation{Theoretical Astrophysics,  IAAT, University of T\"ubingen, T\"ubingen 72076, Germany}

\author{Stoytcho S. Yazadjiev}
\email{yazad@phys.uni-sofia.bg}
\affiliation{Theoretical Astrophysics, IAAT, University of T\"ubingen, T\"ubingen 72076, Germany}
\affiliation{Department of Theoretical Physics, Faculty of Physics, Sofia University, Sofia 1164, Bulgaria}
\affiliation{Institute of Mathematics and Informatics, 	Bulgarian Academy of Sciences, 	Acad. G. Bonchev St. 8, Sofia 1113, Bulgaria}

\date{\today}
%
\begin{abstract}
The aim of this paper is to study the stability of soliton-like static solutions via non-linear simulations in the context of a special class of massive  tensor-multi-scalar-theories of gravity whose target space metric admits Killing field(s) with a periodic flow. We focused on the case with two scalar fields and maximally symmetric target space metric\R{, as} the simplest configuration where solitonic solutions can exist. In the limit of zero curvature of the target space $\kappa=0$\R{,} these solutions reduce to the standard \R{boson stars}, while for $\kappa \ne 0$ significant deviations can be observed, both qualitative and quantitative. \R{By evolving these solitonic solutions in time,} we show that they are stable for low values of the central scalar field $\psi_c$ while instability kicks in with the increase of $\psi_c$. \R{Specifically, in the stable region,} the models oscillate with a characteristic frequency related to the fundamental mode\R{. Such} frequency tends to zero with the approach of the unstable models and eventually becomes imaginary when the solitonic solutions \R{lose} stability.  As expected from the study of the equilibrium models, the change of stability occurs exactly at the maximum mass point\R{, which} was checked numerically with a very good accuracy. 
\end{abstract}
\maketitle

\section{Introduction}

In modern physics, Einstein's theory of General Relativity (GR) represents the official theory of gravity: it provides a quantitative description of gravitational phenomena and passes all present experimental tests. Nevertheless, the study of alternative theories of gravity is supported by different reasons \cite{Will_2014,Berti_2015,Barack:2018yly}. The first motivation is related to the numerous attempts to come up with a consistent quantum theory of the gravity through GR. Since no viable way to incorporate GR in the formalism of Quantum Field Theory has been found, a shortcoming of the theories is expected to be observed in the high energy regime. Motivation also comes from theories trying to unify all the interactions, such as higher dimensional gravity, Kaluza-Klein theory, etc., predicting the existence of scalar fields that are as well mediators of the gravitational interaction. Another issue concerning GR is related to the dark {energy and dark matter} problem, which can be solved by a modification of GR, {as an alternative} to the introduction of an exotic energy-matter component of the universe \cite{Lue_2004}. In the present paper, we will focus on compact objects made entirely by a condensation of gravitational scalars, providing, therefore, possible dark matter models derived from an extension of GR.

Tensor-multi-scalar theories (TMST) represent a class of theories of gravity which provide a natural extension to GR \cite{Damour,TMS_theories}. Unlike the simpler tensor-scalar theories of gravity \cite{bergmann_1968, wagoner_1970}, {TMSTs} are defined by $\mathcal{N}>1$ scalar fields non-minimally coupled to the Ricci scalar in the action and taking value in a coordinate patch of a target-space manifold supplemented with a metric \cite{Damour}. The richer geometrical structure of these theories can affect the dynamics and give more freedom to the parameters that define the particular theory; some parameters are, indeed, left unconstrained by weak-field observations \cite{TMS_theories}.

We focus on a sub-class of theories in which the target space is characterized by the presence of Killing fields with a periodic flow. It has been shown \cite{Yazadjiev_2019} that such sub-class of theories exhibit static soliton-like solutions for the vacuum field equations and the results were later generalized to the case of rotation \cite{Collodel:2020gyp} and mixed soliton-fermion objects \cite{Doneva:2019krb}. These configurations are associated with hypothetical solitonic compact objects formed entirely by a condensation of the gravitational scalars, non-interacting directly with the electromagnetic field, and, thus, dark in nature. In the case when the target space is {flat} these objects coincide with the well known {boson stars (BSs)} \cite{Schunck:2003kk,Liebling:2012fv}, but the non-trivial structure of the target space can lead to qualitative and qantitative differences as shown in \cite{Yazadjiev_2019,Collodel:2020gyp}. The interest in such objects is connected to the fact that their mass can range in huge interval -- from {typical neutron star masses up to the masses of supermassive compact objects in the galactic centers}. Therefore, they provide very interesting astrophysical implication such as black hole (BH) mimickers and a possible solution of the dark matter problem.

In a previous article \cite{Yazadjiev_2019}, the existence of the solitons in TMSTs is numerically investigated in the context of the particular sub-class of {TMSTs} mentioned before. Nevertheless, a further analysis regarding the stability of the solutions has still to be made.  {Specifically, the existence of hypothetical stable and unstable branches}  can be tested, by analogy with classical general relativistic self-gravitating configurations.

The gravitational stability of {BSs has already been} discussed in different previous works. The first approach consisted {of} the study of the time evolution of small perturbations around an equilibrium configuration in the context of linear stability
analysis \cite{Gleiser}. {Such analysis was carried out through perturbations that conserve mass and charge (see \cite{Gleiser1988ih} for the ground state analysis and \cite{JETZER1989447} for the excited states) and resulted in the identification of a change of stability corresponding to the point of maximum mass.} More generally, for perturbations that do not conserve particle number, excited states turned out to be unstable {under} all conditions and they decay to the corresponding ground state \cite{LEE1989477}. The results were {numerically confirmed by non-linear analysis of finite perturbations (see \cite{PhysRevD.42.384} for the zero-node solutions and \cite{Balakrishna_1998} for the cases with self-interaction and excited states)}. The dynamical fate of unstable BS solutions was numerically analysed in \cite{guzman2019dynamical}, resulting in two possible scenarios: unstable bounded solutions that collapse to form BHs and unstable unbounded solutions that disperse to infinity.

In the present paper, we will focus on more general soliton solutions predicted by the TMSTs, that can be quite different from BSs. The aim of the work is to investigate the stability along sequences of such solitonic solutions by performing a nonlinear numerical evolution {in the simplified case of spherical symmetry}.

In Section \ref{the_general_theory} we discuss the general theoretical framework focusing on the definition of the particular sub-class of theories. In Section \ref{the_field_equations} we show all the equations and the boundary conditions related to the static solutions and their evolution in time that {were} implemented in the simulations. In Section \ref{numerical_results} we show the results of the simulations, presenting the expected behaviour of unstable solutions.

\section{The general theory} \label{the_general_theory}

In the Einstein frame the general action of the {TMSTs} of gravity \cite{Damour} writes
\begin{align} \label{action_minimal_coupling}
S[{\varphi}^a,g_{\mu\nu},\Psi_i] & = \frac{1}{16 \pi G_*} \int d^4x \sqrt{-g} \left(R-2\gamma_{ab}(\varphi) g^{\mu\nu} \partial_\mu \varphi^a \partial_\nu \varphi^b - 4 V(\varphi) \right)  \nonumber \\
& + S_m[a^2(\varphi)g_{\mu\nu},\Psi_i]
\end{align}
where $\varphi^a$ ($a = 1,...,\mathcal{N}$) are the scalar fields, $g_{\mu\nu}$ is the Einstein frame metric, $\Psi_i$ are the matter fields, $S_m$ is the matter action, $G_*$ is the bare gravitational constant, $R$ is the Ricci scalar with respect to $g_{\mu\nu}$. The function $\gamma_{ab}(\varphi)$ takes the role of positively definite metric supplementing the $\mathcal{N}$-dimensional Riemannian (target) manifold $\mathcal{E}_\mathcal{N}$, whose coordinate patch is where the scalar fields $\varphi^a$ take value. The function $a(\varphi)$ represents the conformal factor relating the Einstein and the Jordan metrics through the identity $\tilde{g}_{\mu\nu}=a^2(\varphi)g_{\mu\nu}$, where $\tilde{g}_{\mu\nu}$ is the {physical} Jordan frame metric{. Unlike $g_{\mu\nu}$, the metric $\tilde{g}_{\mu\nu}$ is regarded as the ``physical'' gravitational field,} since particles follow its geodesics \cite{sotiriou_liberati_faraoni_2008}{ ; nevertheless, in the present article we will focus on the Einstein frame, since the action (\ref{action_minimal_coupling}) is, by definition, in the more-practical Hilbert-Einstein form}. Both $V(\varphi)$ and $a(\varphi)$ can be seen as scalar fields defined on the target space $\mathcal{E}_\mathcal{N}$ and they specify the particular {TMST}, together with $\gamma_{ab}(\varphi)$.

Following \cite{Yazadjiev_2019}, we will focus on the  $\mathcal{N} = 2$ case {and we choose the so-called isothermal coordinates, as} they are guaranteed in $2$-dimensional manifolds and they are characterized by a metric conformal to the Euclidean one:
\begin{equation} \label{isothermal_coordinates_metric}
\gamma_{ab}(\varphi) = \Omega^2(\varphi) \delta_{ab}
\end{equation}
for a particular $\Omega^2(\varphi)$. {Thus,} the final action (\ref{action_minimal_coupling}) depends exclusively on the functions $\Omega(\varphi)$, $V(\varphi)$ and $a(\varphi)$.

Following the results of \cite{Yazadjiev_2019}, we look for a class of {TMSTs} which exhibit static soliton-like solutions for the field equations in consequence of the following choices: the metric $\gamma_{ab}(\varphi) $ admits a Killing field $K^a$ with a periodic flow and $a(\varphi)$ and $V(\varphi)$ are invariant under the flow of the Killing field $K^a$, i.e. $ \mathcal{L}_{K} V(\varphi) = K^a \partial_a V (\varphi) = 0 $ and $ \mathcal{L}_{K} a(\varphi) = K^a \partial_a a (\varphi) = 0 $.

The existence of a Killing vector suggests the presence of a conserved current in the Einstein frame
\begin{equation} \label{charge_continuity}
J^\mu = -\frac{1}{4 \pi} g^{\mu \nu} K_a \partial_\nu \varphi^a
\end{equation}
such that $\nabla_\mu J^\mu = 0$, with $\nabla$ as the covariant derivative with respect to the Einstein frame metric $g_{\mu\nu}$, and a Jordan frame conserved current
\begin{equation}\label{current}
\tilde{J}^\mu = a^{-4}(\varphi) J^\mu = \frac{1}{4 \pi a^2(\varphi)} \tilde{g}^{\mu \nu} K_a \partial_\nu \varphi^a
\end{equation}
such that $ \tilde{\nabla}_\mu \tilde{J}^\mu = 0$.   Both of them provide a conservative charge (i.e. $ \partial_0 Q = 0 $)
\begin{equation} \label{charge}
Q = \int_{space} J^0 dV = \int_{space} \tilde{J}^0 d\tilde{V}
\end{equation}
where the volume element $ dV  = \sqrt{-g_{00}} \sqrt{h} d^3x $  is taken from the space-like hypersurfaces with constant $ x^0 $ and $h$ is the determinant of the metric $h_{\mu\nu} $ induced on such hypersurfaces.

As in \cite{Yazadjiev_2019} we will focus on the simplest case of $\mathcal{N} = 2$. In the isothermal coordinate system, the Killing field with periodic orbits can be explicitly given by the following coordinates:
\begin{equation}
K^a(\varphi) = - \epsilon^a{}_b \varphi^b
\end{equation}
with $\epsilon^a{}_b = - \epsilon^b{}_a$ and $ \epsilon^1{}_2 = 1$. In order for $K^a$ to be a Killing vector for the metric $ \gamma_{ab}(\varphi) $ (i.e. $ \mathcal{L}_{K}\gamma_{ab}(\varphi) = 0 $), $ \Omega(\varphi) $ must depend on $ \varphi = (\varphi^1, \varphi^2) $ through $ \psi^2 = (\varphi^1)^2 + (\varphi^2)^2 $. The same applies to $ V(\varphi) $ and $ a(\varphi) $ in order for the conditions $ \mathcal{L}_{K} V(\varphi) = 0 $ and $ \mathcal{L}_{K} a(\varphi) = 0 $ to be satisfied.

Other choices that we have to make are about the {form of the free functions} in the action (\ref{action_minimal_coupling}). Following results from \cite{Yazadjiev_2019}, we chose in our simulations the following functions
\begin{subequations}
\begin{equation} \label{Omega}
\Omega(\psi) = \frac{1}{1+\frac{\kappa}{4}\psi^2},
\end{equation}
\begin{equation} \label{V}
V(\psi) = \frac{1}{2} m_s \psi^2 + \frac{1}{4} \lambda_{(4)} \psi^4,
\end{equation}
\begin{equation} \label{a}
a(\psi) = \exp \left( \frac{1}{2} \beta \psi^2 \right).
\end{equation}
\end{subequations}
{Equation (\ref{Omega}) defines a metric $\gamma_{ab}(\psi) $} for a maximally symmetric 2-dimensional manifold, with spherical ($ \kappa > 0 $), hyperbolic ($ \kappa < 0 $) or flat ($ \kappa = 0 $) geometry. Equation (\ref{V}) represents a potential with $m_s$ as mass term and $\lambda_{(4)}$ as parameter for the 4-self-interaction of the scalar fields. As in \cite{Yazadjiev_2019} we will work with $\beta = -6$.

\section{The field equations} \label{the_field_equations}

The field equations resulting from the action (\ref{action_minimal_coupling}) are the following:
\begin{subequations}\label{field_equations}
\begin{equation}\label{Einstein_equations}
G_{\mu \nu} = \gamma_{a b}(\varphi) \left( 2 \nabla_\mu \varphi^a \nabla_\nu \varphi^b - g_{\mu \nu} \nabla_\rho \varphi^a \nabla^\rho \varphi^b \right) - 2 V(\varphi) g_{\mu \nu} + 8 \pi G_* T_{\mu \nu}),
\end{equation}
\begin{equation}\label{multi-scalar_field_equations}
\nabla_\mu \nabla^\mu \varphi^a = -\gamma^a{}_{bc}(\varphi^d) g^{\mu \nu} \nabla_\mu \varphi^b  \nabla_\nu \varphi^c + \gamma^{a b}(\varphi) \left( \frac{\partial V(\varphi)}{\partial \varphi^b} - 4 \pi \alpha_b(\varphi) G_* T \right),
\end{equation}
\end{subequations}
where $G_{\mu\nu}$ is the Einstein tensor related to the Einstein frame metric $g_{\mu\nu}$, $T_{\mu \nu} = a^2(\varphi^a)\tilde{T}_{\mu\nu}$ is the energy-momentum tensor of the matter fields in the Einstein frame, $T$ is its trace, {$\alpha_a$ are defined as} $ \alpha_a = \partial_a \ln a $ and $\gamma^a{}_{bc}(\varphi^d)$ are the Christoffel symbols related to the metric $\gamma_{ab}(\varphi^c) $.

The energy-momentum conservation equation in the Einstein frame can be derived from (\ref{Einstein_equations}) using the contracted Bianchi identities:
\begin{equation}\label{energy_momentum_conservation_Einstein}
\nabla_\nu T_\mu {}^\nu = \nabla_\mu \varphi^a \alpha_a(\varphi) T_\nu {}^\nu
\end{equation}

As in \cite{Yazadjiev_2019} we will focus on spherically symmetric and asymptotic flat space-times. In such space-time, we can adopt a coordinate system $(t,r,\theta,\phi)$, such that the metric $g_{\mu\nu}$ takes the following form

\begin{equation} \label{metric_Einstein}
ds^2 = - N^2(t,r) dt^2 + A^2(t,r) dr^2 + r^2(d\theta^2 + \sin^2\theta d\phi^2).
\end{equation} 
We also define $ \Phi $ and $ m $ from the relations $N = e^\Phi$ and $A = \sqrt{\frac{1}{1-\frac{2m}{r}}}$.

{By adopting radial coordinates system and the void hypothesis (i.e. $T_{\mu\nu}=0$), the field equations (\ref{field_equations})} read explicitly:
\begin{subequations} \label{equations_of_motion_eucl}
\begin{equation} \label{m_prime_eucl}
\partial_r m = r^2 \left( \frac{1}{2} \Xi + V(\psi) \right),
\end{equation}
\begin{equation}  \label{Phi_prime_eucl}
\partial_r \Phi = r A^2 \left( \frac{m}{r^3} + \frac{1}{2} \Xi -  V(\psi) \right) ,
\end{equation}
\begin{equation} \label{m_dot_eucl}
\partial_t m =r^2 \frac{N}{A} \varphi_t{}^a \varphi_r{}_a,
\end{equation}
\begin{align} \label{phi_dot_eucl}
&\partial_t \varphi_t^a = \frac{N}{A} \left( \partial_r \varphi_r^a + \left( \frac{2}{r} + \partial_r \Phi \right) \varphi_r{}^a \right) - \frac{1}{r} A^2 (\partial_t m) \varphi_t{}^a - 2 N \Omega^{-2}(\psi) \frac{dV(\psi)}{d\psi^2} \varphi^a   \nonumber\\
& + 2 N \Omega^{-2}(\psi) \left( \frac{d \ln \Omega(\psi)}{d \psi^2} \left( (\varphi_t{}^b \varphi_t{}_b - \varphi_r{}^b \varphi_r{}_b) \varphi^a - 2 (\varphi_t{}^b \varphi_b \varphi_t{}^a - \varphi_r{}^b \varphi_b \varphi_r{}^a) \right) \right)
\end{align}
\end{subequations}
with $\varphi_t{}^a = \frac{1}{N} \partial_t \varphi^a $, $\varphi_r{}^a = \frac{1}{A} \partial_r \varphi^a $, and $\Xi = \varphi_t{}^a \varphi_t{}_a + \varphi_r{}^a \varphi_r{}_a$.

Soliton-like static configurations can be set up by {a harmonically time-dependent scalar field} \cite{Yazadjiev_2019}:
\begin{equation}\label{stationary_static_condition}
\varphi(r,t) = (\psi(r) \cos (\omega t), \psi(r) \sin(\omega t)).
\end{equation}
Such condition implies that the effective energy-momentum tensor, defined as the right hand side of equation (\ref{Einstein_equations}), is time-independent and the same applies to the geometry of space-time. With condition (\ref{static_equations}), equations (\ref{equations_of_motion_eucl}) read
\begin{subequations} \label{static_equations}
\begin{eqnarray} \label{m_prime_static}
\partial_r m = r^2 \left( \frac{1}{2} \Omega^2(\psi) \left( \psi_r{}^2 + \omega^2 N^{-2} \psi^2 \right) + V(\psi) \right),
\end{eqnarray}
\begin{equation}  \label{nu_prime_static}
\partial_r \Phi = r A^2 \left( \frac{m}{r^3} + \frac{1}{2} \Omega^2(\psi) \left( \psi_r{}^2 + \omega^2 N^{-2} \psi^2 \right) -  V(\psi) \right),
\end{equation}
\begin{align} \label{Psi_prime_prime_static}
 \partial_r \psi_r = & - \left( \frac{2}{r} + \partial_r  \Phi  \right) \psi_r \nonumber\\
& - A \psi \left( \omega^2 N^{-2} - 2 \Omega^{-2}(\psi) \frac{dV(\psi)}{d\psi^2} + 2 \frac{d \ln \Omega(\psi)}{d \psi^2} ( \omega^2 N^{-2}\psi^2+\psi_r{}^2) \right),
\end{align}
\begin{equation}
\partial_r \psi = A \psi_r,
\end{equation}
\end{subequations}
resulting into a set of coupled differential equations to be solved under the conditions of asymptomatically flatness of the space-time (i.e. $\Phi(r=\infty) = 0$), no singularities in $ r=0 $ (i.e. $m(r=0) = 0$), regularity at the center (i.e. $\psi_r(r=0) = 0$) and an arbitrary choice of central value of the scalar field (i.e. $\psi(r=0) = \psi_c$). Equations (\ref{static_equations}) with such boundary conditions define an eigenvalue problem for $\omega$ that can be numerically solved. In this paper we are interested {in} zero-nodes solutions, since they are expected to be the only ones that can provide a stable branch \cite{Balakrishna_1998}.

\section{Numerical results} \label{numerical_results}

Once the parameters of the theory $m_s$, $\kappa$ and $\lambda_{(4)}$ are fixed, any static solution can be univocally defined by the choice of $\psi_c$. {An estimate of the their stability can be deduced from the study of the ADM mass $M$ and the scalar charge $Q$ as functions of $\psi_c$. Nevertheless, a rigorous test can be made only} by studying the time evolution of a compact object after imposing a small initial perturbation. 

During the whole simulation we can keep track of the evolution of some relevant quantities. Specifically the maximum of $A(r)$ and {the minimum of} $N(r)$ are expected to, respectively, diverge and converge to zero, {when an event horizon appears}. Moreover, we can show the results in units of $m_s$, or, equivalently, using adimensional quantities (e.g.: $\lambda_{(4)} m_s^{-2}$, $R m_s$, $M m_s$, $Q m_s^2$). In this way $m_s$ becomes the parameter that fixes the scale of the system.

We expect a qualitative and quantitative change in the stability branch for different values of $\lambda = \lambda_{(4)} m_s^{-2}$ and $\kappa$. In \cite{Yazadjiev_2019} it has been shown that an increase {in} $\lambda$ and $\kappa$ leads to an increase {in} the maximum mass $M$ and charge $Q$ that solutions with different $\psi_c$ can reach and a decrease {in} the maximal point. {The result is that the range of central values of $\psi$ for which stable solutions exists shrinks}.

\begin{figure}[h]
\includegraphics[scale=0.46]{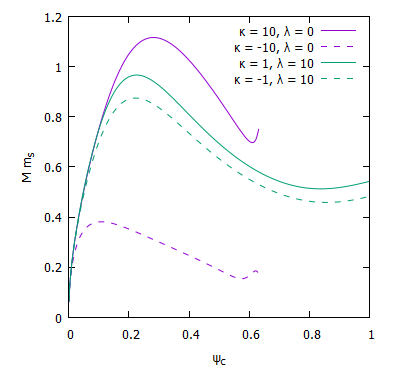}
\includegraphics[scale=0.46]{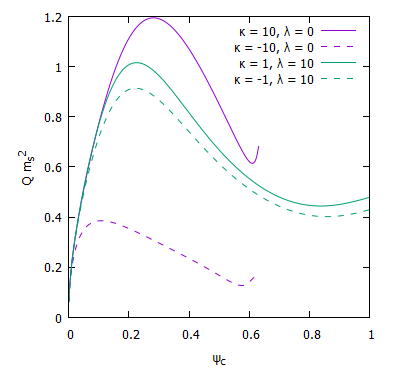}
\caption{The normalized mass {$Mm_s$} (left panel) and the normalized conserved charge $Qm_s^2$ (right panel) as functions of the central value of the scalar field $\psi_c$ with different combinations of the parameters $(\kappa=\pm 10,\lambda=0)$ and $(\kappa=\pm 1,\lambda=10)$} \label{mass_charge}
\end{figure}

Stable solutions with $\kappa=0$ are BSs made of an exotic complex field and already known in literature, as we have mentioned before. For this reason, we will focus on some cases with $\kappa \neq 0$. In order to have relevant differences from the BS case, we choose the following four possible combinations of the two parameters: $(\kappa=\pm 10,\lambda=0)$ and $(\kappa=\pm 1,\lambda=10)$. The corresponding normalized mass $M m_s$ and the normalized conserved charge $Q m_s^2$ {are presented in Figure \ref{mass_charge} for sequences of solitonic solutions}. The branches of solutions exhibit a first maximum mass and charge at the following maximal points: $\psi_c^{(max)} = 0.282 \pm 0.001 $ for $(\kappa= 10,\lambda=0)$,  $\psi_c^{(max)} = 0.109 \pm 0.001 $ for $(\kappa= 10,\lambda=0)$, $\psi_c^{(max)} = 0.226 \pm 0.001 $ for $(\kappa= 1,\lambda=10)$ and $\psi_c^{(max)} = 0.222 \pm 0.001 $ for $(\kappa= -1,\lambda=10)$.

Stable configurations are expected to be the ones with $\psi_c<\psi_c^{(max)}$. Solutions in the unstable branch and close to the maximal point are instead expected to exhibit instability from small perturbations and collapse into BHs. In order to test the stability of generic configurations, we have numerically simulated the time evolution of solutions with $\psi_c \lesssim \psi_c^{(max)}$ and  $\psi_c \gtrsim \psi_c^{(max)}$.

With the above obtained solutions as a background, equations (\ref{Phi_prime_eucl}), (\ref{m_dot_eucl}), and (\ref{phi_dot_eucl}) are used in order to perform the time evolution. This is done through finite differencing. The variables $m$, $\Phi$ and $\varphi^a$, are defined on grid points, while $\varphi^a_t$ and $\varphi^a_r$ are defined halfway between grid points. Starting from an initial condition $t=0$, already set up by the static conditions, we calculate step by step the value of $m$, $\varphi$, for every $ t = \bar{t} + k $, knowing the value of the variables in $ t=\bar{t} $ using equations (\ref{m_dot_eucl}) and $\partial_t\varphi^a=N \varphi^a_t$. In a second step the constraint (\ref{Phi_prime_eucl}) is used in order to find $\Phi(t = \bar{t} + k)$: starting from the numerical infinity, where $\Phi (r=\infty) \approx 0$, we evaluate the variable $\Phi$ by moving backward with respect to $r$, until we reach $r=0$. Finally, the value of $\varphi^a_t$ between $t = \bar{t} + k$ and $t = \bar{t} + 2k$ is evaluated using equation (\ref{phi_dot_eucl}) and knowing the configuration at $t = \bar{t} + k$.

We use, as boundary condition for the scalar field at the numerical infinity, {the one} of an outgoing spherical wave on a Schwarzschild background \cite{PhysRevD.42.384}
\begin{equation}
\partial_t \varphi_t^a = - \partial_t \varphi_r^a -  2 N  \frac{dV(\psi)}{d\psi^2} \varphi^a - \frac{1}{r A} \varphi^a ,
\end{equation}
which gives rise to the dispersion relation for a massive Klein-Gordon field in the asymptotically flat region of the space-time, up to the first order of $m_s / f$ (where $f$ is the frequency of an outgoing wave). A massive Klein-Gordon field with mass $m_s$ can approximate a field solution of equation (\ref{phi_dot_eucl}) in the asymptotically flat region up to the first order of $\varphi^a$ and its derivatives.

\begin{figure}[h]
\begin{center}
\includegraphics[scale=0.6]{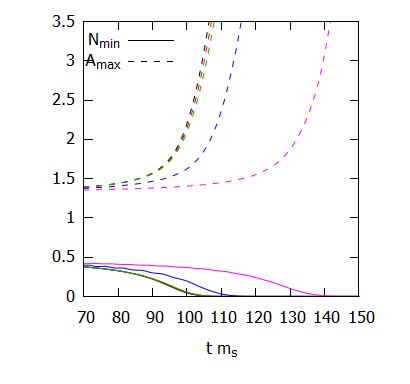}
\end{center}\caption{{Time evolution of the maximum of the metric function $A$ and minimum of the metric function $N$} is shown for the unstable solution with $\kappa=10$, $\lambda=0$ and $\psi_c=0.285$ and with different values of the radial resolutions of the grid: $\Delta r = 1 m_s^{-1}$ (black), $\Delta r = 2 m_s^{-1}$ (red), $\Delta r = 3 m_s^{-1}$ (green), $\Delta r = 4 m_s^{-1}$ (blue), $\Delta r = 5 m_s^{-1}$ (magenta).} \label{convergency}
\end{figure}

We can test the stability of the solutions by numerically {evolving} {equations \eqref{equations_of_motion_eucl}}. Indeed, small deviation from the analytical solution are always introduced due to the numerical errors generated in every step of the simulated evolution. Moreover, in order to amplify the effects of the perturbations and to have a quicker collapse, we work with low temporal precision {-- specifically with $\Delta t \sim 10^{-2} T$ (where $T = 2 \pi  / \omega$) --} for the first $\sim 50$ steps{. Such procedure} effectively introduces additional random perturbation of the background solution. 

The configurations with $\psi_c \gtrsim \psi_c^{(max)}$ are expected to collapse into BHs as result of the instability; an example is given by Figure \ref{convergency}. The temporal evolution of such configuration is shown with different radial resolutions of the grid{, with the aim of providing a convergence} test for the simulations: in the continuum limit, the solutions appear to converge into a single one, ideally unaffected by numerical errors. The difference in the collapse time is due the different numerical error, resulting in a different perturbation applied to each solution; in the continuum limit, such difference appears to vanish, proving the robustness of the code.

\begin{figure}[h]
\includegraphics[scale=0.47]{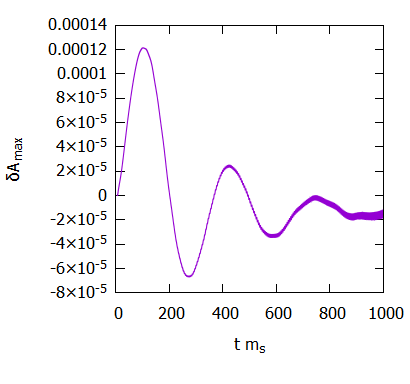}
\includegraphics[scale=0.47]{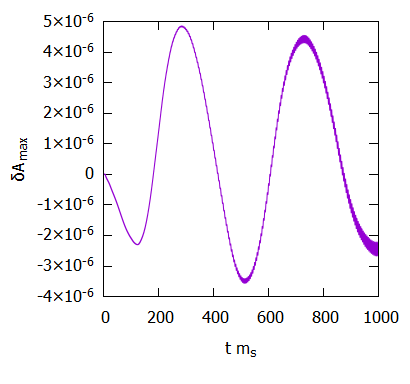}
\caption{{Time evolution for the perturbation of the maximum value of {the metric function} $A$ ($\delta A_{max}(t) = A_{max}(t) - A_{max}(0)$) for the stable solution with $\kappa=10$, $\lambda=0$ and $\psi_c=0.272$ (left panel) and $\psi_c=0.278$ (right panel)}} \label{fundamental_mode}
\end{figure}

{Solutions with $\psi_c \lesssim \psi_c^{(max)}$ have been tested and they remain stable, even after imposing a perturbation introduced by the numerical errors -- the perturbed stable solution oscillates for a while and then settle down close to the equilibrium stable solution. The signal in this case is a mixture of different modes in the context of linear analysis. Nevertheless, we focus only on the fundamental mode, since it is the one responsible for instability \cite{Gleiser1988ih}. Specifically, the critical point $\psi_c^{max}$ is associated to the fundamental mode frequency turning into zero and becoming imaginary in the unstable branch. Such property has been numerically tested, with different values of $\psi_c$ approaching to $\psi_c^{max}$ from below. In Figure \ref{fundamental_mode}, some stable time evolutions are shown, revealing a decreasing value of the fundamental mode frequency for increasing $\psi_c$ and thus with the approach of the critical point. Moreover, the stability properties, such as the turning of the fundamental mode frequency to zero and the time evolution in the instability region have been checked to be robust against changes of resolution.}


The results of the time evolution for some unstable solutions are, instead, summarized in Figure \ref{kappa=1,lambda=10}, \ref{kappa=-1,lambda=10}, \ref{kappa=10,lambda=0}, \ref{kappa=-10,lambda=0}. In all of these cases, the configurations collapse into BHs.\ This is suggested by the metric {function} $N$ that collapses to zero in a region expected to be covered by a horizon and the metric function $A$ that starts to diverge in correspondence of the coordinate singularity. It is also possible to notice how the BH region, whose inside is characterized by $N \ll 1$ and whose border by $A \gg 1$, is delimited by a radial size close to the Schwarzschild radius{. What happens is that} almost all the scalar field condensate {ends up} inside the BH, while the rest {is} radiated away. The temporal coordinate in which the BH appears varies with different initial perturbations; specifically, as mentioned before, an amplification for such initial perturbation {is} artificially introduced in order to have a quicker collapse.

We have performed evolution of a large number of background solitonic solutions and the results show that the relevant property emerging from the simulations is that the stability condition (i.e.: $\psi_c < \psi_c^{(max)}$) is independent of the theory parameters $\kappa$ and $\lambda$. Moreover, for particular choices of such parameters, the critical point has been tested with a precision of the order of $\Delta \psi_c^{(max)} / \psi_c^{(max)} \approx 10^{-2}$.

\begin{figure}[h]
\includegraphics[scale=0.52]{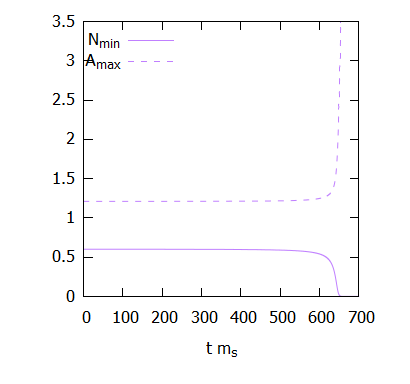}
\includegraphics[scale=0.52]{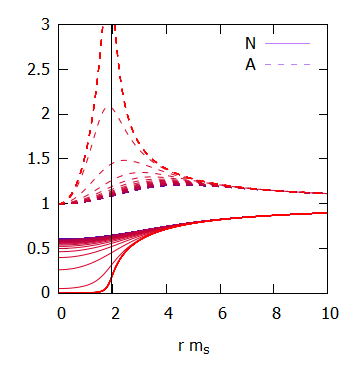}
\caption{
Time evolution for the unstable solution with $\kappa=1$, $\lambda=10$ and $\psi_c=0.227$; on the left figure, {the maximum and the minimum of the metric functions $A$ and $N$ correspondingly} is shown, suggesting the formation of a horizon when $A$ diverges and $N$ converge to zero; on the right figure, the radial profile of the metric functions $A$ and $N$ is shown for different timesteps, with step $\Delta t = 2 \pi  / \omega$, from blue (${tm_s} = 0$) to red (${tm_s} = {100 \cdot 2 \pi} / \omega = 762$); the black line represents the Schwarzschild radius of the initial configuration.} \label{kappa=1,lambda=10}
\end{figure}

\begin{figure}[h]
\includegraphics[scale=0.52]{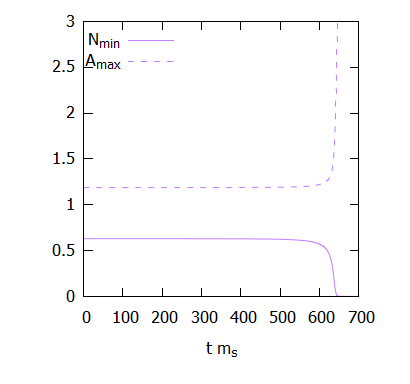}
\includegraphics[scale=0.52]{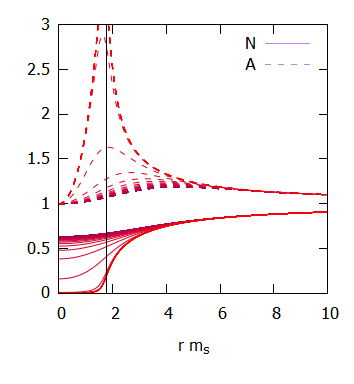}
\caption{
Time evolution for the unstable solution with $\kappa=-1$, $\lambda=10$ and $\psi_c=0.223$; on the left figure, {the maximum and the minimum of the metric functions $A$ and $N$ correspondingly} is shown, suggesting the formation of a horizon when $A$ diverges and $N$ converge to zero; on the right figure, the radial profile of the metric functions $A$ and $N$ is shown for different timesteps, with step $\Delta t = 2 \pi  / \omega$, from blue (${tm_s} = 0$) to red (${tm_s} = {100 \cdot 2 \pi} / \omega = 750$); the black line represents the Schwarzschild radius of the initial configuration.} \label{kappa=-1,lambda=10}
\end{figure}

\begin{figure}[h]
\includegraphics[scale=0.52]{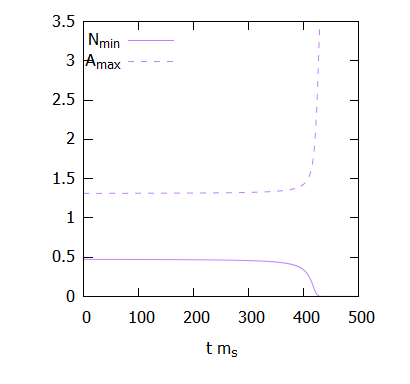}
\includegraphics[scale=0.52]{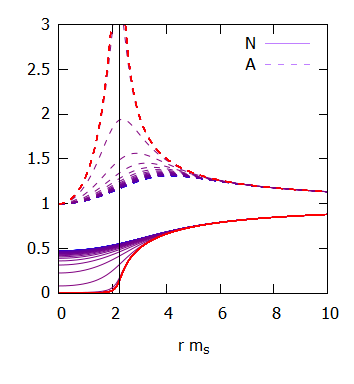}
\caption{
Time evolution for the unstable solution with $\kappa=10$, $\lambda=0$ and $\psi_c=0.285$; on the left figure, {the maximum and the minimum of the metric functions $A$ and $N$ correspondingly} is shown, suggesting the formation of a horizon when $A$ diverges and $N$ converge to zero; on the right figure, the radial profile of the metric functions $A$ and $N$ is shown for different timesteps, with step $\Delta t = 2 \pi  / \omega$, from blue (${tm_s} = 0$) to red (${tm_s} = {100 \cdot 2 \pi} / \omega = 823$); the black line represents the Schwarzschild radius of the initial configuration.} \label{kappa=10,lambda=0}
\end{figure}

\begin{figure}[h]
\includegraphics[scale=0.52]{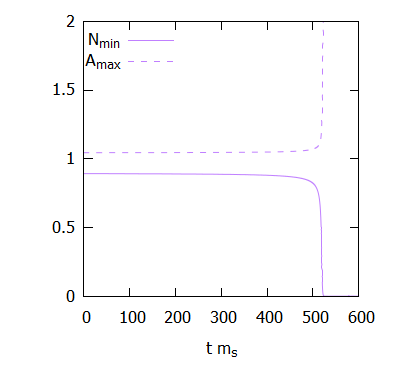}
\includegraphics[scale=0.52]{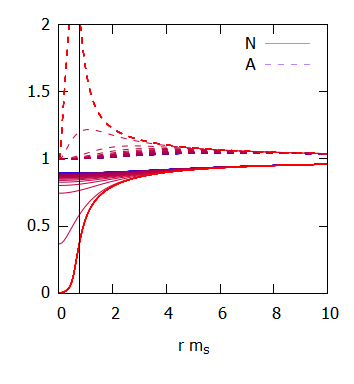}
\caption{
Time evolution for the unstable solution with $\kappa=-10$, $\lambda=0$ and $\psi_c=0.110$; on the left figure, {the maximum and the minimum of the metric functions $A$ and $N$ correspondingly} is shown, suggesting the formation of a horizon when $A$ diverges and $N$ converge to zero; on the right figure, the radial profile of the metric functions $A$ and $N$ is shown for different timesteps, with step $\Delta t = 2 \pi  / \omega$, from blue (${tm_s} = 0$) to red (${tm_s} = {100 \cdot 2 \pi} / \omega = 665$); the black line represents the Schwarzschild radius of the initial configuration.} \label{kappa=-10,lambda=0}
\end{figure}

\section{Conclusion} \label{conclusion}

In the present paper we have studied the stability of soliton-like static solutions in the context of a special class of massive {TMSTs} of gravity whose target space metric admits Killing field(s) with a periodic flow. We focused on the case with two scalar fields and maximally symmetric target space metric that is the simplest configuration where solitonic solutions can exist. In the limit of zero curvature of the target space $\kappa=0$ these solutions reduce to the standard BSs, while for $\kappa \ne 0$ significant deviations can be observed, both qualitative and quantitative.    

For this purpose the relevant field equations describing the time evolution of the static spherically symmetric solutions  are derived, assuming that the variables depend only on the radial coordinate and the time. Even though this is clearly a simplification, we should note that we do not perform any linearlization procedure, thus keeping a more general form of the problem beyond the standard analysis of the linearized radial perturbations. For the solution of the resulting partial differential equations, a numerical code is implemented. The evolution of a series of soliton solutions along the obtained sequences of background solutions is studied. As expected from the examination of the equilibrium configurations, models before the maximum of the mass are stable and they relax to the equilibrium solutions after imposing a small perturbation. After the maximum of the mass, though, the models quickly collapse to a BH. The change of stability happens exactly at the maximum mass point for all combinations of the parameters.  

Let us comment on the possible extensions of these results. As a matter of fact the developed code has greater capabilities, namely the study of the radial collapse and the emitted gravitational radiation. This will be of particular observational interest, since the end configuration will be a BH without scalar hair and thus the scalar field has to be radiated away fully before the actual BH formation.  Such a work in underway and will be reported in a future publication.

\section*{Acknowledgement}

RF thanks the hospitality of the Theoretical Astrophysics group in {T\"ubingen} where a significant part of the work has been done. DD acknowledges financial support via an Emmy Noether Research Group funded by the German Research Foundation (DFG) under grant no. DO 1771/1-1. SY would like to thank the University of T\"ubingen for the financial support.  SY acknowledge financial support by the Bulgarian NSF Grand KP-06-H28/7. Networking support by the COST actions CA16104 and CA16214 is gratefully acknowledged.

\bibliography{bibliography} 
\bibliographystyle{ieeetr}

\end{document}